\documentclass[1p,times]{elsarticle}

\usepackage{amsfonts}
\usepackage{amsthm}
\usepackage{amssymb}
\usepackage{amsmath}
\usepackage{graphicx}
\usepackage{color}
\usepackage{xcolor}
\usepackage{bm}
\usepackage{times,pslatex}

\usepackage{float}
\usepackage{enumitem}

\usepackage{float}

\RequirePackage{verbatim}

\RequirePackage{color}
\definecolor{todo}{rgb}{1,0,0}
\definecolor{answer}{rgb}{0,0,1}
\definecolor{new}{rgb}{1,0,1}
\definecolor{conditional}{rgb}{0,1,0}
\definecolor{e-mail}{rgb}{0,.40,.80}
\definecolor{reference}{rgb}{.20,.60,.22}
\definecolor{mrnumber}{rgb}{.80,.40,0}
\definecolor{citation}{rgb}{0,.40,.80}

\usepackage[colorlinks, linkcolor=reference,
 citecolor=citation,
         urlcolor=e-mail]{hyperref}

\usepackage{bbm}

\newtheorem{theorem}{Theorem}
\newtheorem{lemma}{Lemma}

\newtheorem{proposition}{Proposition}

\theoremstyle{remark}
\newtheorem{example}{Example}

\newtheorem{ourdefinition}{Definition}
\newtheorem{remark}{Remark}

\usepackage{todonotes}
\usepackage{algorithm}
\usepackage{algorithmic}

\newcommand{\CC}{\mathbb{C}}

\newcommand{\NN}{\mathbb{N}}
\newcommand{\KK}{\mathbb{K}}
\newcommand{\ksmall}{\mathbb{F}}
\newcommand{\LL}{\mathbb{L}}
\newcommand{\sys}{\mathcal{F}}
\newcommand{\Va}{\mathbb{V}}
\newcommand{\Pa}{\mathcal{P}}
\newcommand{\LD}{\mathcal{L}}

\DeclareMathOperator{\numer}{numer}
\DeclareMathOperator{\denom}{denom}

\begin{document}
\begin{frontmatter}

\title{Algorithm for globally identifiable reparametrizations of ODEs\tnoteref{t1,t2}}
 \tnotetext[t1]{
     The final version of this article is part of the volume titled
     ``Computational Algebra and Geometry:
     A special issue in memory and honor of Agnes Szanto''.}
     \tnotetext[t2]{ The paper is dedicated to the memory of Agnes Szanto, to whom we are grateful for her open-mindness, inspiration, and mentorship.}

\author{Sebastian Falkensteiner}
\ead{sebastian.falkensteiner@mis.mpg.de}
\address{Max Planck Institute for Mathematics in the Sciences, Leipzig, Germany}
\author{Alexey Ovchinnikov}
\ead{aovchinnikov@qc.cuny.edu}
\address{Department of Mathematics, CUNY Queens College and Ph.D. Programs in Mathematics and Computer Science, CUNY Graduate Center, New York, USA}
\author{J. Rafael Sendra}
\ead{jrafael.sendra@cunef.edu}
\address{Department of Quantitative Methods, CUNEF-University, Spain}

\begin{abstract}
Structural global parameter identifiability indicates whether one can determine a parameter's value in an ODE model from given inputs and outputs. If a given model has parameters for which there is exactly one value, such parameters are called  globally identifiable. Given an ODE model involving not globally identifiable parameters, first we transform the system into one with locally identifiable parameters. As a main contribution of this paper, then we present a procedure for replacing, if possible, the ODE model with an equivalent one that has globally identifiable parameters. We first derive this as an algorithm for one-dimensional ODE models and then reuse this approach for higher-dimensional models.
\end{abstract}


\end{frontmatter}
\section{Introduction}
Estimating values of parameters in ODE systems from time series data is a central problem in modeling. The structure of the ODE system could imply that more than one or even infinitely many parameter values fit the data. In these cases, we say that parameters are not globally (but locally) identifiable and are non-identifiable, respectively. 
Algorithms for checking local identifiability can, for instance, be found in~\cite{chis2011structural,hermann1977nonlinear,sedoglavic2001probabilistic}. 
Approaches to global identifiability are described, for example, in~\cite{chics2011genssi,hong2020global,pohjanpalo1978system,saccomani2008daisy,walter1982global}.
If not all of the parameters of a model are globally identifiable,~\cite{ollivier1990probleme} presents a method to find rational functions of the parameters that are identifiable from the corresponding input-output (IO) equations, and~\cite{denis2004equivalence,dong2023differential,meshkat2009algorithm,Ovchinnikov2021} from the given ODE system.

 ODE systems with parameters that are not globally identifiable are particularly challenging for further parameter estimation. In particular, optimization-based parameter estimation methods are very frequently used in practice. However, even if a parameter is locally but not globally identifiable, the errors in estimates obtained by optimization based methods~\cite{SBtoolbox,DifferentialEquations.jl-2017,rackauckas2020universal,AMIGO2} can easily be much higher than for globally identifiable models, see the locally identifiable Biohydrogenation, Mammillary 4, and SEIR models in the tables in~\cite{par_est_robust} (in the present paper, we will have SEIR as one of our examples).  This is because one typically obtains only one solution
for the parameter values using optimizers even if there are multiple solutions fitting
into a physically meaningful range, and the value found by the optimizer could be one of those but rather far from the original true value. Finding a globally identifiable reparametrization would give the user a guarantee that the
solution returned by such a parameter estimation algorithm is unique.

Repairing non-identifiability of ODE systems has been a topic in several projects. 
For an algorithmic method to simplify ODE systems by differential elimination, i.e., reduce the number of equations, differential indeterminates etc., we refer to~\cite{boulier2007differential}. 
In~\cite{ilmer2022more}, algebraically independent non-identifiable parameters are found and eliminated in the given ODE system. 
Linear ODE models are treated in~\cite{baaijens2016existence,meshkat2014identifiable}. 
For non-linear models, we refer to~\cite{chappell1998procedure,evans2000extensions} in the case of replacing the model with one where the parameters are locally identifiable. For example,  ad-hoc methods for finding locally and globally identifiable reparametrizations are discussed in~\cite{denis2004equivalence,meshkat2009algorithm}. 
An algorithm for actually computing the new model in the identifiable parameters is presented in~\cite{OPPS2023}, which imposes very restrictive conditions on the support of the new models and guarantees only local identifiability of the reparametrized model even though globally identifiable reparametrizations might be available. Another reparametrization algorithm, based on restricting the support of Lie derivatives of the reparametrized model, is given in~\cite{MOS2023}.

In this paper, we study the problem of finding, for a given ODE model involving  not globally identifiable parameters in the form of a rational or polynomial realizations, an alternative ODE model in which the parameters are globally identifiable and the ODE system remains rational or polynomial, respectively. 
The main contribution of our paper is to derive an algorithm for addressing this problem. 
As a preparation step, we present an algorithm to transform the given ODE model into another model with locally identifiable parameters; the algorithm is similar to~\cite[Algorithm 1]{ilmer2022more}, presented in the framework of the current paper. Then we apply a method trying to transform this model into an ODE model with just globally identifiable parameters.
Since many ODE models in practice are of polynomial form, we focus on polynomial realizations and the so-called realizability problem in terms of polynomial realizations. 

The question of reconstructing a dynamical system in the state space from a given set of IO-equations is known as the realizability problem and is widely studied in control theory~\cite{ jakubczyk1980minimal,schaft1986classic,sontag1993rational}. 
The structure of the problem depends significantly on  which class the functions are sought in. 
Rational functions were used in~\cite{falkensteiner2023real,pavlov2022realizing}. 
In this paper, we mainly address the case of polynomial functions.

The structure of the paper is as follows. 
In Section~\ref{sec-prel}, we recall several necessary results from differential algebra, introduce the notion of realizations, identifiability of parameters and observability, and relate them to rational and polynomial parametrizations. In addition, we recall the technique, based on hypercircles and ultraquadrics, to simplify the coefficients of a rational parametrization.
In Section~\ref{sec-polynomialrealizations}, we study the problem of finding polynomial realizations of first-order IO-equations. In Lemma~\ref{lem-help1}, we show that either every realization depends polynomially on the input variable or none of them. Using this result, the existence of polynomial realizations for first-order IO-equations can be decided, see Algorithm~\ref{alg-polynomialrealization}. Moreover, if there exists such a polynomial realization, then it can always be chosen to be observable, as it is shown in Theorem~\ref{thm-polynomialobservable}.
The approach used for these results and the algorithm can be extended to higher order IO-equations to some extent.

Section~\ref{sec-identifiability} is devoted to the problem of replacing a given ODE model whose parameters are locally identifiable with another one with the same IO-equations such that the involved parameters are globally identifiable. The transformation into an ODE model with locally identifiable parametrizations is considered in~\ref{sec:aux}, followed by the main contribution of the paper: the transformation into an ODE model with just globally identifiable parameters. In the case of first-order IO-equations, the Weil's descent variety, and more precisely the witness variety associated to the parametrization, gives a criterion to decide the existence of such a reparametrization (Algorithms~\ref{alg-optimalrealization} and~\ref{alg-optimalpolynomialrealization}). In particular, in Theorem~\ref{thm-optimalrealization}, we show that, if a first-order IO-equation admits a polynomial realization and there is a rational realization with identifiable parameters, then there is also a polynomial realization with identifiable parameters.
This approach can again be generalized to higher-order IO-equations, which we explain in Section~\ref{sec:43} and also illustrate there using  both academic and real-life examples, including an identifiable reparametrization of a locally identifiable SEIR epidemiology model.

\section{Preliminaries}\label{sec-prel}
Let $\CC(\mathbf{c})$ be the complex numbers attached by a set of parameters $\mathbf{c}=(c_1,\ldots,c_k)$. 
In the following, we will use bold letters for vectors and apply operations on them component-wise.
In this paper, we study ODE systems of the form
\begin{equation}\label{eq-realization}
    \Sigma =
    \begin{cases}
    \mathbf{x}' = \mathbf{f}(\mathbf{u},\mathbf{c},\mathbf{x}), \\
    \mathbf{y} = \mathbf{g}(\mathbf{u},\mathbf{c},\mathbf{x})
    \end{cases}
\end{equation}
where $\mathbf{f},\mathbf{g}$ are tuples of rational functions in their arguments, $\mathbf{x}$ is the tuple of state variables, $\mathbf{y}$ the tuple of output variables, $\mathbf{u}$ the tuple of input variables and $\mathbf{c}$ are the parameters.

\subsection{Differential algebra and identifiability}
We give a few basic notions from differential algebra that are later used to define input-output (IO) equations and IO-identifiability.
\begin{ourdefinition}
\begin{enumerate}
\item[]
    \item A {\em differential ring} $(R,')$ is a commutative ring with a derivation $':R\to R$, that is, a map such that, for all $a,b\in R$, $(a+b)' = a' + b'$ and $(ab)' = a' b + a b'$. 
    \item The {\em ring of differential polynomials} in the variables $x_1,\ldots,x_n$ over a field $\KK$ is the ring $\KK[x_j^{(i)}\mid i\geqslant 0,\, 1\leqslant j\leqslant n]$ with a derivation defined on the ring by ${\left(x_j^{(i)}\right)}' := x_j^{(i + 1)}$. This differential ring is denoted by $\KK\{x_1,\ldots,x_n\}$.  Its field of fractions is denoted by $\KK\langle x_1,\ldots,x_n\rangle$.
    \item An ideal $I$ of a differential ring $(R,')$ is called a {\em differential ideal} if, for all $a \in I$, $a'\in I$. For $F\subset R$, the smallest differential ideal containing the set $F$ is denoted by $[F]$.
    \item For an ideal $I$ and element $a$ in a ring $R$, we denote \[I \colon a^\infty = \left\{r \in R \mid \exists \ell\colon a^\ell r \in I\right\}.\] This set is also an ideal in $R$.
    \item Given $\Sigma$ as in~\eqref{eq-realization}, we define the prime differential ideal~\cite[Lemma~3.2]{hong2020global} of $\Sigma$ as \[I_\Sigma=\left[Q\,\mathbf{x}'-Q\,\mathbf{f},Q\,\mathbf{y}-Q\,\mathbf{g}\right]:Q^\infty \subset \CC(\mathbf{c})\{\mathbf{x},\mathbf{y},\mathbf{u}\},\] where $Q$ is the common denominator of $\mathbf{f}$ and $\mathbf{g}$.
\end{enumerate}
\end{ourdefinition}

In the following, we will set $\KK = \CC(\mathbf{c})$ and work with subfields of $\KK$.

\begin{ourdefinition}[IO-identifiability]\label{def:IO-ident}
 Given an ODE system $\Sigma$, the smallest field $\ksmall$ such that $\CC \subset \ksmall \subset \CC({\mathbf c})$ and $I_\Sigma \cap \CC({\mathbf c})\{\mathbf{y},\mathbf{u}\}$ is generated as an ideal by $I_\Sigma \cap \ksmall\{\mathbf{y},\mathbf{u}\}$ is called the \emph{field of IO-identifiable functions}  of $\Sigma$.
We call a rational function $h \in \CC({\mathbf c})$ \emph{IO-identifiable} if $h \in \ksmall$. We also call $h \in \CC({\mathbf c})$ \emph{locally IO-identifiable} if $h$ is in the algebraic closure of the field $\ksmall$.
\end{ourdefinition}

\begin{ourdefinition}[IO-equations]
For a fixed differential ranking $>$ on $(\mathbf{y},\mathbf{u})$, the set of \textit{input-output equations (IO-equations)} of the system $\Sigma$ from~\eqref{eq-realization} is the characteristic presentation of
\begin{equation}\label{eq-IOEquations}
I_{\Sigma} \cap \KK\{\mathbf{y},\mathbf{u}\}.
\end{equation}
See \cite[Section~5.2]{ident-compare} for more details, including showing that the field $\ksmall$ of identifiable functions is equal to the field generated by the coefficients of IO-equations.
\end{ourdefinition}

\begin{example}
Consider the differential system
\[
\Sigma =\begin{cases}
x' = ax +b\\
y = x
\end{cases}
\]
Here $\KK = \CC(a,b)$.
To compute IO-equations of $\Sigma$, we consider the differential elimination ranking $x>y$, and the corresponding characteristic presentation is
\[C = \{x'-ax - b, y'-ax - b\}.\]
By a property of characteristic presentations with respect to elimination rankings, \[\{y'-ay - b\}=C\cap \KK\{y\}\] is a characteristic presentation of $I_\Sigma\cap \KK\{y\}$, and so $y'-ay-b = 0$ is the IO-equation of $\Sigma$ and $\ksmall = \KK=\CC(a,b)$ is the field of IO-identifiable functions.
\end{example}

\subsection{State-space realizations of IO-equations and parametrizations}
\begin{ourdefinition}[Realizability]\label{def:realization}
Starting from a system of differential equations $\sys \subset \KK\{\mathbf{y},\mathbf{u}\}$, an ODE-system $\Sigma$ of the form~\eqref{eq-realization} such that $\sys$  is a set of IO-equations of $\Sigma$
is called a \textit{realization} of $\sys$.
Moreover, if there exists a realization $\Sigma$ of $\sys$, then we call $\sys$ \textit{realizable}.
\end{ourdefinition}

Note that there can be no realization or several realizations for $\sys$. 
As we will see later, a necessary condition on the realizability of $\sys$ is that the implicitly defined variety admits a rational parametrization.

In the following, we will consider for simplicity just a single input variable $u$ and note that the case of several input variables can be considered in a similar way.

Let us now formalize the discussion from Example~\ref{ex:LV}.
Given $\sys \subset \KK\langle u\rangle\{\mathbf{y}\}$ in the output variables $y_1,\ldots,y_r$ of order $n_i$ in $y_i$, we define the \textit{corresponding algebraic variety} with coordinates $y_1,\ldots, y_1^{(n_1)},\ldots,y_r,\ldots,y_r^{(n_r)}$ as the zero set
\[ \Va(\sys) = \{ P \in \overline{\KK\langle u\rangle}^n \mid F(P)=0 \text{ for every } F \in \sys \} \]
where $n=\sum_{i=1}^r n_i+r$ and where, for a field $\mathbb{L}$, $\overline{\mathbb{L}}$ denotes the algebraic closure of $\mathbb{L}$.

\begin{example}[Lotka-Volterra]\label{ex:LV}
Consider the system
\begin{equation}
\label{eq:LV}
\Sigma=
\begin{cases}
x_1' = ax_1 -bx_1x_2,\\
x_2'= -cx_2 + dx_1x_2,\\
y = x_1
\end{cases}
\end{equation}
with two state variables $\mathbf{x} = (x_1,x_2)$, four parameters $\mathbf{c}=(a,b,c,d)$, and one output $\mathbf{y} = y$. We have $\KK = \CC(a,b,c,d)$. A calculation shows that the IO-equation is
\begin{equation}
\label{eq:LVIO}
F = yy'' - y'^2 -dy^2y' + cyy' + ady^3 - acy^2 .
\end{equation}
Therefore, the field of IO-identifiable functions is
$\CC(d,c,ad,ac) = \CC(a,c,d)$. Equation~\eqref{eq:LVIO} defines an irreducible affine variety (surface) $\Va(F)$ in the coordinates $y, y', y''$ over the field $\CC(a,c,d)$. In the terminology of Definition~\ref{def:realization}, $\Sigma$ is a realization of $F$.
Let us compute the Lie derivatives of the $y$-variable with respect to the vector field defined by $\Sigma$ to understand the connection between $V$ and $\Sigma$:
\begin{equation}\label{eq:LieLV}
\begin{aligned}
y &= x_1, \\
y' &= -bx_1x_2 + ax_1,\\
y'' &= -bdx_1^2x_2 + b^2x_1x_2^2  + (bc-2ab)x_1x_2 + a^2x_1
\end{aligned}
\end{equation}
These equations define a polynomial parametrization of $\Va(F)$ over the ground field for $\Sigma$, which is $\CC(a,b,c,d)$.
\end{example}

Let us formally describe the observations from Example~\ref{ex:LV}.

\begin{ourdefinition}[Parametrizations]
A \textit{(uni-)rational parametrization} of a variety $\Va(\sys)$ is a dominant rational map from some affine space to $\Va(\sys)$. Equivalently, this is a point $\Pa \in \overline{\KK{\langle u\rangle}}(x_1,\ldots,x_d)^n$, with some new transcendental parameters $\mathbf{x}=(x_1,\ldots,x_d)$, vanishing on $\sys$ such that the Jacobian matrix of $\Pa$ w.r.t. $\mathbf{x}$, denoted by $\mathcal{J}(\Pa)$, has full rank, that is, $\mathcal{J}(\Pa)$ has a $d\times d$ submatrix $\mathcal{M}$ whose determinant is a non-zero rational function in $x_1,\ldots,x_d$.
\end{ourdefinition}

\begin{ourdefinition}
A rational parametrization $\Pa$ is called \textit{polynomial} if each component of the parametrization is a polynomial, that is, $\Pa \in \overline{\KK\langle u\rangle}[x_1,\ldots,x_d]^n$.
\end{ourdefinition}

\begin{ourdefinition}
A rational parametrization is called \textit{proper} (or birational) if it induces a birational map from the affine space to the algebraic variety almost everywhere  in the Zariski topology.
\end{ourdefinition}

\begin{ourdefinition}
For given $P \in \LL(u,\mathbf{x}), \mathbf{f} \in \LL(u,\mathbf{x})^d$, where $\LL \subseteq \overline{\KK}$, we define the Lie derivative of $P$ w.r.t. $\mathbf{f}$ as (see~\eqref{eq:LieLV} for an example)
\[ \LD_{\mathbf{f}}(P) = \sum_{i=1}^d \partial_{x_i}P \cdot f_i + \sum_{i \ge 0}\partial_{u^{(i)}}P \cdot u^{(i+1)} \]
and denote the second sum by $D_u(P)$. 
\end{ourdefinition}

For a given realization $\Sigma$ as in~\eqref{eq-realization}, we can find a parametrization of $\Va(\sys)$ as
\begin{equation}\label{eq-parametrization}
    \Pa = (g_1,\LD_{\mathbf{f}}(g_1),\ldots,\LD_{\mathbf{f}}^{n_1}(g_1),\ldots,g_r,\LD_{\mathbf{f}}(g_r),\ldots,\LD_{\mathbf{f}}^{n_r}(g_r)) \in \LL\langle u\rangle(\mathbf{x})^n
\end{equation}
where $\LD_{\mathbf{f}}^i$ denotes the $i$-th iteration of $\LD_{\mathbf{f}}$. 
Let us note that each component of $\Pa$ may dependent on derivatives of $u$ up to the order of the iteration of $\LD_{\mathbf{f}}$. We call $\Pa$ in~\eqref{eq-parametrization} the parametrization corresponding to the realization $\Sigma$ in~\eqref{eq-realization}.

For notational purposes, in the following we assume that the invertible $d \times d$ submatrix of $\mathcal{J}(P)$ is of the form
\begin{equation}\label{eq-invertiblesubmatrix}
\mathcal{M} = \begin{pmatrix}
\partial_{x_1}P_{1,0} & \cdots & \partial_{x_d}P_{1,0} \\
\vdots & & \vdots \\
\partial_{x_1}P_{1,m_1} & \cdots & \partial_{x_d}P_{1,m_1} \\
\vdots & & \vdots \\
\partial_{x_1}P_{r,0} & \cdots & \partial_{x_d}P_{r,0} \\
\vdots & & \vdots \\
\partial_{x_1}P_{r,m_r} & \cdots & \partial_{x_d}P_{r,m_r}
\end{pmatrix},
\end{equation}
where $0 \le m_i < n_i$.

\begin{theorem}[{cf.~\cite{Forsman} if no inputs and~\cite{falkensteiner2023real} if one output}]
Let \[\Pa=(P_{1,0},\ldots,P_{1,n_1},\ldots,P_{r,0},\ldots,P_{r,n_r}) \in \LL\langle u\rangle(\mathbf{x})^n\] be a rational parametrization of $\Va(\sys)$ with the invertible $d \times d$ submatrix $\mathcal{M}$ as in~\eqref{eq-invertiblesubmatrix}. 
Then $\Pa$ defines a realization of $\sys$ if and only if
\begin{enumerate}
    \item for all $i \in \{1,\ldots,r\}$ it holds that $P_{i,0} \in \LL\langle u \rangle (\mathbf{x})$; and
    \item \begin{equation}\label{eq-param}
    \mathbf{z} = \mathcal{M}^{-1} \cdot (P_{1,1}-D_u(P_{1,0}),\ldots,P_{r,m_r+1}-D_u(P_{1,m_r}))^T\in \LL\langle u \rangle (\mathbf{x})^d.
    \end{equation}
\end{enumerate}
In the affirmative case, the realization is \[\begin{cases}\mathbf{x}'=\mathbf{z},\\y_i=P_{i,0}.
\end{cases}\]
\end{theorem}

In this paper, we are mainly interested in realizations that are polynomial. 
\begin{ourdefinition}[Polynomial realizations]
A realization~\eqref{eq-realization} is called \textit{polynomial} if the right-hand sides $\mathbf{f}, \mathbf{g}$ are polynomial in $\mathbf{u}$ and $\mathbf{x}$.
\end{ourdefinition}
The parametrization $\Pa$ corresponding to a polynomial realization is polynomial as well. 
Polynomial parametrizations are studied in~\cite{sendra2008rational,sendra2001optimal}. The latter paper has  an algorithm for computing polynomial parametrizations of curves and  an optimal coefficient field for it, but there is no general algorithm for computing polynomial parametrizations of varieties of dimension two or bigger (for a particular type of surfaces, an algorithm is provided in~\cite{perez2020computing}).

Additionally to polynomial realizations and polynomial parametrizations, the case where the parametrization is proper is of particular interest. 
This corresponds to the case where all states $\mathbf{x}$ are (globally) observable, that is~\cite[Proposition 3.4, Theorem 3.16]{hong2020global}, $$\overline{\KK}\langle u\rangle(g_1,\LD_\mathbf{f}(g_1),\ldots,\LD_\mathbf{f}^{n_r}(g_r)) = \overline{\KK}\langle u\rangle(\mathbf{x}).$$
\begin{ourdefinition}[Observable realizations]
We call a realization \textit{observable} if the corresponding parametrization is proper.
\end{ourdefinition}

\begin{example}
Let $F=y'$, so the variety $\Va(F)$ is the line given by $y'=0$ in the ambient affine 2-space with coordinates $y,y'$. The zero $\Pa = (x^2, 0)$ is a polynomial paramerization because $\mathcal{J}(\Pa) = (2x, 0)$ has rank one. However, $\Pa$ is not proper because, for all $a \ne 0$, there are $b$, $c$ such that $b\ne c$ and $b^2 = c^2  = a$. 
Therefore, the realization 
\[\Sigma = \begin{cases} x'= 2x,\\ y = x^2\end{cases}\] of $F=y'$ is not observable. 
The realization \[\Sigma_0 = \begin{cases} x'=1, \\ y=x\end{cases},\] however, is observable.
\end{example}

\subsection{Simplifying coefficients of parametrizations}\label{sec:simplcoeff}
The IO-identifiability of the rational function components of a rational  parametrization is related to their coefficients, and hence we will study  whether we can reparametrize the parametrization with simpler coefficients. In this section, we recall how to approach this question by means of the witness variety of the parametrization. The ideas described below come from~\cite{andradas1999base,andradas2009simplification,recio2010generalizing}. We refer to these papers for further details.

Let $\mathbb{L}$ be a field of characteristic zero and $\mathbb{L}(\alpha)$ a finite algebraic extension of $\mathbb{L}$ of degree $m$ given by a primitive element $\alpha$. Let $\Pa(\mathbf{x})\in \mathbb{L}(\alpha)(\mathbf{x})^n$, say $\mathbf{x}=(x_1,\ldots,x_d)$, be a proper rational parametrization of an algebraic variety $V$ in the $n$--dimensional affine space over the algebraic closure of $\mathbb{L}(\alpha)$. Without loss of generality, we assume that $\Pa$ is written as 
\[ \Pa=\left( \dfrac{p_1}{q}, \ldots, \dfrac{p_n}{q}\right) \]
where $\gcd(p_1,\ldots,p_n,q)=1$.
We are interested in deciding, and actually computing, whether $\Pa$ can be birationally reparametrized into a parametrization of $V$ with coefficients in $\mathbb{L}$. For this, we consider the basis $\mathcal{B}:=\{1,\alpha,\ldots,\alpha^{m-1}\}$ of $\mathbb{L}(\alpha)$ as an $\mathbb{L}$-vector space, and we formally write each $x_i$ in the $\mathcal{B}$-basis as 
\[ x_i=z_{i,0}+z_{i,1}\alpha+\cdots+ z_{i,m-1} \alpha^{m-1},\]
where $z_{i,j}$ are new variables; let $\mathbf{z}:=(z_{1,0},\ldots,z_{n,m-1})$. We replace each $x_i$ in $\Pa$ by its representation in the $\mathcal{B}$--basis. After writing the new rational functions in the $\mathcal{B}$--basis, $\Pa$ 
can be expressed as 
\[ \Pa=\left( \sum_{j=0}^{m-1}\dfrac{H_{1,j}(\mathbf{z})}{\delta(\mathbf{z})} \,\alpha^j, 
\ldots, \sum_{j=0}^{m-1}\dfrac{H_{n,j}(\mathbf{z})}{\delta(\mathbf{z})}\, \alpha^j \right)
\]
where $H_{i,j},\delta\in \mathbb{L}[\mathbf{z}]$ are coprime. 

\begin{ourdefinition}\label{def:witness}
With the previous notation, let 
\begin{enumerate}
\item $\mathrm{Y}$ be the variety over $\overline{\mathbb{L}}$ defined by the polynomials 
$\{ H_{i,j}(\mathbf{z})\}_{1\leq i \leq n, 1\leq j \leq m-1}$, 
\item $\Delta$ be the variety over $\overline{\mathbb{L}}$ defined by the polynomial $\delta(\mathbf{z})$.
\end{enumerate}
We define the {\em $\alpha$-witness variety} associated to $\Pa$ as the Zariski closure of $\mathrm{Y}\setminus \Delta.$
\end{ourdefinition}

The possibility of parametrizing $\Pa$ over $\mathbb{L}$ can be deduced from the  witness  variety using the theory of hypercircles (see \cite{recio2010generalizing} for the case of dimension one) and ultraquadrics (see \cite{andradas2009simplification} for higher dimension). Nevertheless, here, we shortcut the theory recalling   slightly weaker statements. 

\begin{proposition}[{cf.~\cite[Theorem 17]{andradas2009simplification}}]\label{proposition-witness-variety}
Let $V$ be properly parametrizable over $\mathbb{L}$. Then the  witness variety   has a component that is parametrizable over $\mathbb{L}$ with a parametrization given by the coordinates w.r.t. the basis $\mathcal{B}$ of an $\mathbb{L}$--automorphim of $\mathbb{L}(\mathbf{z})$. Furthermore, if $(\phi_0,\ldots,\phi_m)$ is such a parametrization,  then \[\Pa(\phi_0+\phi_1 \alpha +\cdots + \phi_m \alpha^m)\] is a proper parametrization of $V$ over $\mathbb{L}$.
\end{proposition}

\begin{proposition}[{cf.~\cite[Theorem~15]{andradas2009simplification}}]\label{prop:hypercircles}   Let $\dim(V ) = 1$. $V$ can be parametrized over 
$\mathbb{L}$ iff the witness
 variety has a one-dimensional component parametrizable over $\mathbb{L}$. In this case, if
$(\phi_0,\ldots, \phi_m)$ is a proper parametrization over $\mathbb{L}$ of that component, then
\[ \Pa(\phi_0 + \phi_1 \alpha + \cdots+ \phi_m\alpha^m) \]
is a proper parametrization of $V$ over $\mathbb{L}$.
\end{proposition}

The case of polynomial parametrizations of curves has a special treatment (see  \cite[Theorem~3]{sendra2001optimal} and  \cite[Theorem~2.6]{recio2010generalizing}).

\begin{proposition}\label{prop:lines} Let $\dim(V ) = 1$ and  $\Pa$ polynomial. $V$ can be polynomially parametrized over 
$\mathbb{L}$ iff the witness
 variety has a component being a line definable over $\mathbb{L}$. In this case, if
$(\phi_0,\ldots, \phi_m)$ is a proper polynomial parametrization over $\mathbb{L}$ of the line, then
\[ \Pa(\phi_0 + \phi_1 \alpha + \cdots+ \phi_m\alpha^m) \]
is a  proper polynomial parametrization of $V$ over $\mathbb{L}$.
\end{proposition}

Summarizing, one may proceed as follows:

\begin{enumerate}
\item Compute the witness  variety of $\Pa$.
\item Determine its irreducible components.
\item For each component of dimension $\dim(V)$, check whether it is properly parametrizable over $\mathbb{L}$. If so, let $(\phi_0,\ldots,\phi_m)$ be a proper parametrization over $\mathbb{L}$ of the correspondent component.
\item Check whether $\Pa(\phi_0+\phi_1 \alpha +\cdots + \phi_m \alpha^m)$ is over $\mathbb{L}$.
\end{enumerate}

\section{Polynomial realizations}\label{sec-polynomialrealizations}
In this section, we present our theory and algorithm for finding observable polynomial realizations of IO-equations. We begin with showing how our approach works for first-order IO-equations in Section~\ref{sec:1stord}. Even though this might look too restrictive for practical purposes, it contains enough subtleties to help us build an approach for higher-order cases, which we discuss in Section~\ref{sec:higherorder}.

\subsection{First-order IO-equations}\label{sec:1stord}
In this subsection, we consider dynamical systems in a single output variable $y$ and one state variable $x$. This corresponds to a single IO-equation $F \in \KK[u,u',y,y']$ of order one. 
We note that the order of $F$ in $u$ is bounded by the order of $F$ in $y$ (see~\cite[Proposition 2.3]{falkensteiner2023real}). 
(Rational) realizations of first-order IO-equations have been studied in~\cite{pavlov2022realizing,falkensteiner2023real}. 
We will being by establishing several technical results: about polynomal parametrizations of curves (Section~\ref{sec:polpar}) and polynomial realizations of IO-equations (Section~\ref{sec:polreal}).

\subsubsection{Technical preparation: polynomial parametrizations of curves}\label{sec:polpar}
Let us first recall some results on polynomial paramerizations of curves. 
There exists a polynomial parametrization $\Pa$ of $\Va(F)$, over an algebraically closed field $\overline{\KK(u,u')}$ if and only if there exists a proper polynomial parametrization. 
Moreover, every polynomial parametrization $\mathcal{Q}$ of $\Va(F)$ can be found as a reparametrization $\Pa(s)=\mathcal{Q}(x)$ with $s \in \overline{\KK(u,u')}[x]^{2}$. 
Thus, it remains to decide whether there is a polynomial parametrization fulfilling~\eqref{eq-param}. 
For details on the theory and actual computation of polynomial parametrizations of algebraic curves we refer to~\cite{sendra2001optimal}.

\begin{lemma}\label{lem-coprimality}
Let $A, B \in \overline{\KK}[x]$ be non-constant coprime polynomials and let $s \in \overline{\KK}[z] \setminus \overline{\KK}$. 
Then $A(s), B(s) \in \overline{\KK}[z]$ are coprime.
\end{lemma}
\begin{proof}
Assume that $A(s)$ and $B(s)$ have a common factor. 
Since $A(s), B(s)$ are non-constant, there is a common root $\alpha \in \overline{\KK}$. Then $s(\alpha)$ is a root of both $A$ and $B$ in contradiction to their coprimality. 
\end{proof}

\begin{proposition}\label{prop:condition_polynomialrealizability}
Let $F \in \KK[u,u',y,y']$. 
Then $F$ is polynomially realizable if and only if $\Va(F)$ admits a polynomial parametrization $\Pa \in \overline{\KK}[u,x] \times \overline{\KK}(u,u')[x]$ such that~\eqref{eq-param} is in $\overline{\KK}[u,x]$. 
\end{proposition}
\begin{proof}
Let $F$ be polynomially realizable. 
Then the corresponding parametrization is polynomial and belongs to $\overline{\KK}[u,x] \times \overline{\KK}(u,u')[x]$ such that~\eqref{eq-param} is in $\overline{\KK}[u,x]$.

For the converse direction, let \[\Pa=(P_0,P_1) \in \overline{\KK}[u,x] \times \overline{\KK}(u,u')[x]\] be a polynomial parametrization of $\Va(F)$ satisfying~\eqref{eq-param}. 
Then \[\begin{cases}x'=\dfrac{P_1-\tfrac{\partial P_0}{\partial u}\,u'}{\tfrac{\partial P_0}{\partial x}},\\  y=P_0
\end{cases}
\]
is independent of $u'$ and, by assumption, polynomial in $u$ and $x$ such that it defines a polynomial realization of $F$.
\end{proof}

The next lemma gives a criterion on whether rational functions can be transformed into polynomials and is a straight-forward generalization of~\cite[Theorem~3]{manocha1991rational}.

\begin{lemma}
\label{lem-polynomialitycheck}
Let $f_1,\ldots,f_n \in \KK(x)$. 
There is $s \in \overline{\KK}(z) \setminus \overline{\KK}$ such that $f_i(s)$ are polynomial for all $i=0,\ldots,n$ if and only if the common denominator of the $f_i$ is of the form $a \cdot (x-b)^m$ for some $a,b \in \KK$, $m \in \NN$.
\end{lemma}

\subsubsection{Technical results: polynomial realizations}\label{sec:polreal}

\begin{lemma}\label{lem-help1}
Let $F \in \KK[u,u',y,y']$ be a realizable IO-equation. 
Then either every realization of $F$ depends polynomially on $u$ or none of them.
\end{lemma}
\begin{proof}
By~\cite[Theorem 3.9]{falkensteiner2023real}, there exists an observable realization $\Sigma$ of $F$.
By~\cite[Proposition~2.8]{falkensteiner2023real}, every (rational) realization $\Sigma_0$ of $F$ is found by a reparametrization $s \in \overline{\KK}(z)$ of $\Sigma$. 
Thus, if in a denominator of $\Sigma$, say $q(u,x)$,  $u$ occurs effectively, then this is the case also for $q(u,s)\partial_z(s)$ since $s$ is independent of $u$ and non-constant. 
Thus, $\Sigma_0$ depends non-polynomially on $u$. 
If \[\Sigma=\begin{cases} x'=f(x,u),\\ y=g(x,u) \end{cases}\] depends polynomially on $u$, then this is also the case for \[\Sigma_0 = \begin{cases} z'=\cfrac{f(s,u)}{\partial_z(s)},\\ y=g(s,u).\end{cases}\qedhere\]
\end{proof}

\begin{proposition}\label{prop-everyparametrizationleadstorealization}
Let $F \in \KK[u,u',y,y']$. 
Then either every proper polynomial parametrization $\Pa \in \overline{\KK}[u,x] \times \overline{\KK}(u,u')[x]$ of $\Va(F)$ leads to a polynomial realization of $F$ or none of them. 
Moreover, every polynomial realization is found by a reparametrization $s \in \overline{\KK}[z]$ of $\Pa$, and for an affine-linear $s$, $\Pa(s)$ always corresponds to an observable realization of $F$.
\end{proposition}
\begin{proof}
Let $\Pa \in \overline{\KK}[u,x] \times \overline{\KK}(u,u')[x]$ be a proper parametrization of $\Va(F)$ and assume that a reparametrization $\mathcal{Q}=\Pa(u,s)$ with $s \in \overline{\KK}(z)$ defines a polynomial realization of $F$; note that $s$ has to be independent of $u$ and $u'$~\cite[Theorem~3.10]{falkensteiner2023real}. 
By~\cite[Proposition~2.8]{falkensteiner2023real},~\eqref{eq-param} corresponding to $\Pa$ is in $\overline{\KK}(u,x)$, because otherwise there would not be any rational and thus polynomial realization of $F$. 
Then $\Pa$ defines the (rational) realization
\[\Sigma_{\Pa} = \begin{cases}x'=\dfrac{P_1-\tfrac{\partial P_0}{\partial u}\,u'}{\tfrac{\partial P_0}{\partial x}} =: \frac{A(u,x)}{B(u,x)},\\
\noalign{\vspace*{1mm}}
y=P_0(u,x).
\end{cases}
\]
Assume that $A, B \in \overline{\KK}[u,x]$ are coprime. 
The realization corresponding to $\mathcal{Q}$ is
\[\Sigma_{\mathcal{Q}} = \begin{cases} z'=\dfrac{A(u,s)}{B(u,s)\,\partial_z(s)}, \\ \noalign{\vspace*{1mm}} y=P_0(u,s). 
\end{cases}\]
Since $P_0$ is polynomial in $x$, $s$ has to be polynomial as well such that $P_0(u,s) \in \overline{\KK}[u,z]$. 
By Lemma~\ref{lem-help1}, $B$ is independent of $u$. 
Let us write $\Sigma_{\Pa}$ as
\[\begin{cases} 
x' = a_0(x) + a_1(x)u + \cdots + a_d(x)u^d, \\ y= P_0(u,x),
\end{cases}\]
where $a_0,\ldots,a_d \in \overline{\KK}(x)$ and $P_0 \in \overline{\KK}[u,x]$, and since $\Sigma_{\mathcal{Q}}$ is polynomial, $\frac{a_0(s)}{\partial_z(s)},\ldots,\frac{a_d(s)}{\partial_z(s)}$ are polynomial in $z$. 
By Lemma~\ref{lem-coprimality}, $a_0,\ldots,a_d \in \overline{\KK}[x]$ and $\Sigma_{\Pa}$ is already a polynomial realization of $F$.
Let \[\Sigma = \begin{cases} x'=f(x,u), \\ y=g(x,u) \end{cases}\] be an observable polynomial realization of $F$. 
By~\cite[Proposition~3.4]{falkensteiner2023real}, every (rational) realization of $F$ is obtained by a reparametrization of $\Sigma$. 
In particular, every polynomial realization is found in this way. 
In the case where $s$ is affine-linear, then
\[\begin{cases}
z'=\dfrac{f(s,u)}{\partial_z(s)},\\ 
y=g(u,s)
\end{cases}
\]
is a polynomial realization and again observable; observe that $f(s,u)$ is polynomial and by dividing by $\partial_z(s)=-1/z^2$, the polynomiality is preserved.
\end{proof}

For  an algorithm for finding polynomial parametrizations of a plane curve, see, e.g.~\cite{manocha1991rational,sendra2001optimal}. 
For computing polynomial realizations, we need to restrict to polynomial parametrizations in $\overline{\KK}[u,x] \times \overline{\KK}(u,u')[x]$ (see Proposition~\ref{prop-everyparametrizationleadstorealization}). Finding parametrizations of this particular type is in general algorithmically a difficult task. For some examples, however, such parametrizations can be quickly found.

\begin{example}
Let \[F= -y^4 - 12y^3 + 2y'^3 - 54y^2 - 108y - 81.\] 
Then $\Va(F)$ admits the polynomial parametrization $\Pa=(-x^3/4 - 3, x^4/8)$ and leads to polynomial realization \[
\begin{cases}x'=-\dfrac{x^2}{6},\\ 
\noalign{\vspace*{1mm}}
y=-\dfrac{x^3}{4} - 3.
\end{cases}\]
For every reparamerization $P(s)$ with affine-linear $s \in \CC[x]$, we obtain another polynomial realization of $F$. 
For $\deg_x(s)>1$, however, the realization is not polynomial anymore.
\end{example}

\begin{example}
Let \[F= y'^3 + y^2 - 3y'^2 + 3y' - 1.\]
Then $\Va(F)$ admits the proper polynomial parametrization $\Pa=(x^3,-x^2+1)$ which leads to  non-polynomial realization \[\begin{cases}x'=-\dfrac{x^2 - 1}{3x^2}, \\y=x^3.
\end{cases}
\]
\end{example}

For a general algorithm, we avoid this question by using a different approach as it will be used also in Theorem~\ref{thm-polynomialobservable}:
First, we verify whether $F$ is realizable~\cite[Algorithm 2]{pavlov2022realizing} and use, in the affirmative case, an observable realization~\cite[Algorithm 1]{falkensteiner2023real}. 
The realization might be rational in $x$, but has to be polynomial in $u$ (see Lemma~\ref{lem-help1}). Then we try to make the realization polynomial in $x$ via reparametrization.
\subsubsection{Main theorem and algorithm for first-order equations}
\begin{theorem}\label{thm-polynomialobservable}
Let $F \in \KK[u,u',y,y']$. 
If $F$ admits a polynomial realization, then there is an observable polynomial realization of $F$.
\end{theorem}
\begin{proof}
By~\cite[Theorem 3.9]{falkensteiner2023real}, if $F$ admits a polynomial realization \[\Sigma_{\mathcal{Q}} = \begin{cases}w'=p(w,u),\\ y=q(w,u),\end{cases}\] then there exists an observable (rational) realization
\[\Sigma_{\Pa} = \begin{cases}z'=f(x,u),\\y=g(x,u)\end{cases}\]
of $F$ with corresponding parametrization $\Pa=(P_0,P_1)$. 
Since $F$ admits a polynomial realization, by Lemma~\ref{lem-help1}, the denominators of $f$ and $g$ are independent of $u$. 
Let us write $g$ as
\[ y= g_0(x) + g_1(x)\,u + \cdots +g_e(x)\,u^e \]
where $g_0,\ldots,g_e \in \overline{\KK}(x)$. 
From~\cite[Theorem 3.10]{falkensteiner2023real} we know that the polynomial realization $\Sigma_{\mathcal{Q}}$ is obtained by a reparametrization of $\Sigma_{\Pa}$ with $s \in \overline{\KK}(w)$ and is of the form
$$\Sigma_{\mathcal{Q}} = \begin{cases} w' = \cfrac{f(s,u)}{\partial_w(s)}, \\ y= g_0(s) + g_1(s)\,u + \cdots + g_e(s)\,u^e \end{cases}.$$
By Lemma~\ref{lem-polynomialitycheck}, the common denominator of $g_0,\ldots,g_e$ is of the form $a \cdot (w-b)^m$. 
Let us set $r:=\frac{1+bz}{z}$. 
Then
\[\Sigma_0 := \begin{cases}x'=\cfrac{f(r,u)}{r'} =: \tilde{f}(z,u), \\y=g(r,u) =: \tilde{g}(z,u).
\end{cases}\]
is an observable realization where $\tilde{g}(z,u) \in \overline{\KK}[u,z]$. 
Again by~\cite[Theorem 3.10]{falkensteiner2023real}, $\Sigma_{\mathcal{Q}}$ is obtained from $\Sigma_0$ by a reparametrization with $s \in \overline{\KK}(z)$. In particular, $\tilde{g}(s,u)=q(z,u)$ remains polynomial such that $s \in \overline{\KK}[z]$. 
Let us write $\tilde{f}$ as
\[ \tilde{f} = a_0(z) + a_1(z)\,u + \cdots + a_d(z)\,u^d \]
where $a_0,\ldots,a_d \in \overline{\KK}(z)$. Since $\Sigma_{\mathcal{Q}}$ is polynomial, $\frac{a_0(s)}{\partial_z(s)},\ldots,\frac{a_d(s)}{\partial_z(s)}$ are polynomial in $z$. 
By Lemma~\ref{lem-coprimality}, $a_0,\ldots,a_d \in \overline{\KK}[z]$ and $\Sigma_0$ is already a polynomial realization of $F$.
\end{proof}

Based on Theorem~\ref{thm-polynomialobservable}, one possible approach for deriving a general algorithm for finding polynomial realizations is to follow~\cite[Algorithm 2]{pavlov2022realizing} with some adaptations. 
We, however, directly follow the proof of Theorem~\ref{thm-polynomialobservable} in order to derive the following algorithm.

\begin{algorithm}[H]
\caption{PolynomialRealization FirstOrder}
\label{alg-polynomialrealization}
\begin{algorithmic}[1]
    \REQUIRE An irreducible polynomial $F \in \KK[u,u',y,y']$ over a computable field $\KK$.
    \ENSURE A polynomial realization of $F$ if it exists.
    \STATE Check whether $F$ is (rationally) realizable (e.g. by~\cite[Algorithm 2]{pavlov2022realizing}).
    \STATE In the affirmative case, by~\cite[Algorithm 1]{falkensteiner2023real}, compute an observable (rational) realization\[\begin{cases}x'=f(x,u),\\ y=g(x,u)
    \end{cases}\] of $F$ and let $\Pa=(g,\LD_f(g))$ be the corresponding parametrization.
    \STATE If $u$ effectively occurs in the denominator of $f$ or $g$, then stop. Otherwise, denote by $g_0,\ldots,g_e$ the coefficients of $g$ seen as polynomial in $u$, and let $q$ be its common denominator.
    \STATE If $q$ is constant, return \[\begin{cases}x'=f(x,u),\\ y=g(x,u)
    .
    \end{cases}
    \]
    If $q$ is not of the form $a \cdot (x-b)^m$ for some $a,b \in \overline{\KK}$ with $m$ greater or equal to the degrees of the numerators of $f_i, g_j$, then stop.
    \STATE In the remaining case, set $s=\cfrac{1+bz}{z}$.
    \STATE If $\cfrac{f(s,u)}{\partial_z(s)}$ is polynomial, then return the observable polynomial realization \[\begin{cases}x'=\dfrac{f(s,u)}{\partial_z(s)}, \\ \noalign{\vspace*{1mm}} y=g(s,u).
    \end{cases}\]
\end{algorithmic}
\end{algorithm}

\begin{remark}
One might first check whether $\Va(F)$ admits a polynomial parametrization. If this is not the case, then there cannot exist a polynomial realization of $F$.
\end{remark}

\begin{example}
Let \[F=uy^4 + u^3y' - u^2u'y - y^4.\] 
Then an observable (rational) realization is found by
\[ \begin{cases}x'=\dfrac{u - 1}{x^2}, \\ \noalign{\vspace*{1mm}} y = \dfrac{u}{x}\end{cases} \]
with corresponding proper parametrization \[\Pa = (u/x, -u(u - 1)/x^4 + u'/x).\]
The denominator $q=x^4$ can be dissolved by the reparametrization $s=1/z$. 
This leads to the polynomial realization \[
\begin{cases}z'=(1-u)z^4,\\ y=uz.
\end{cases}
\]
\end{example}

\subsection{Higher order realizations}\label{sec:higherorder}
In this section, we will show how the approach that we developed for first-order equations can be used for higher-order equations. Though we begin with showing a subtlety that the general higher-order case has.
\subsubsection{Additional subtlety}\label{sec:limitation}
In the case of higher-order IO-equations, in general, it is not possible to relate observable polynomial realizations by affine-linear transformations (cf. Proposition~\ref{prop-everyparametrizationleadstorealization}) as demonstrated in the following example.
\begin{example}
Consider $F=y+y'-y''$ and the observable realizations \[\Sigma_1 = \begin{cases} x_1'=x_2,\\ x_2'=x_1+x_2,\\ y=x_1 \end{cases}\] and \[\Sigma_2 = \begin{cases} z_1'=-2z_2^3 - 2z_1z_2 - 2z_2^2 + z_2,\\ z_2'=z_1+z_2+z_2^2,\\ y=z_1+z_2^2 .\end{cases}\]
The corresponding parametrizations $\Pa = (x_1,x_2,x_1+x_2)$ and $\mathcal{Q}=(z_2^2 + z_1, z_2, z_2^2 + z_1 + z_2)$, respectively, are related by the transformation $\Pa(\mathbf{s})=\mathcal{Q}$ where $\mathbf{s}=(z_2^2 + z_1,z_2)$.
\end{example}

\subsubsection{General approach, originating from first-order equations}
We can follow the scheme of Algorithm~\ref{alg-polynomialrealization} also in the case of higher order IO-equations  to try to find for a given rational ODE system \[\Sigma = \begin{cases} \mathbf{x}' = \mathbf{f}(\mathbf{u},\mathbf{x}),\\ \mathbf{y}=\mathbf{g}(\mathbf{u},\mathbf{x}) \end{cases}\] a polynomial realization of the same IO-equation. 
For this, we need to check whether (see~\cite[Prop. 2.8]{falkensteiner2023real})
\[
\Sigma_0 = \begin{cases}
\mathbf{x}' = \mathcal{J}(\mathbf{s})^{-1} \cdot \mathbf{f}(\mathbf{u},\mathbf{s}), \\
\mathbf{y} = \mathbf{g}(\mathbf{u},\mathbf{s})
\end{cases}
\]
has polynomial right hand sides for some reparametrization $\mathbf{s} \in \overline{\KK}(\mathbf{z})^d$. 
In general, however, we do not have an  algorithm for this problem nor is it true that $\Sigma$ can always be chosen observable (see e.g.~\cite[Section 3]{falkensteiner2023real}).

\begin{remark}
Necessary conditions on the given rational ODE system such that it can be reparametrized into a polynomial realization are:
\begin{enumerate}
    \item $\mathbf{g}$ does not have denominators that effectively depend on $\mathbf{u}$ (cf. Lemma~\ref{lem-help1});
    \item\label{item:2} The corresponding variety $\Va(\mathcal{F})$ of the system of IO-equations $\mathcal{F}$ admits a polynomial parametrization.
\end{enumerate}
In~\cite{perez2020computing}, it is described how a proper rational parametrization of certain types of surfaces might be transformed into a proper polynomial parametrization if it exists. 
This approach can be used to answer item~\ref{item:2} above in the case of $d=2$ and a single IO-equation for corresponding algebraic varieties of such type. Moreover, in the affirmative case, reparametrizations leading to polynomial $\mathbf{g}(\mathbf{u},\mathbf{s})$ are computed. 
An adaptation for curves and surfaces in higher-dimensional ambient space could be given.
\end{remark}

For the difficulties of the problem of turning a given non-observable realization $\Sigma$ into an observable realization we refer to~\cite{falkensteiner2023real}.

\begin{example}
Consider the second-order IO-equation \[F = -4u^2y + 4u^2y' - u^2y'' - 6uu'y + 3uu'y' + uu''y - 3u'^2y\] of the ODE system
\[ \Sigma = \begin{cases} x_1' = -\dfrac{u}{x_2^2}-2x_1x_2, \\ x_2' = 2, \\ \noalign{\vspace*{1mm}} y = ux_1. \end{cases} \]
To dissolve the denominator $x_2^2$, we invert the second component. The other expressions must remain polynomial, which can be achieved by the reparametrization $\mathbf{s}=(z_1z_2,1/z_2)$ leading to the polynomial realization
\[ \Sigma_0 = \begin{cases} z_1' = u, \\ z_2' = 2z_2, \\ y = uz_1z_2. \end{cases} \]
\end{example}

\section{Realizations with identifiable parameters}\label{sec-identifiability}

In~\cite{Ovchinnikov2021,dong2023differential} and~\cite[Section~5.2]{ident-compare}, it is discussed how to compute a generating set of IO-identifiable functions $\mathbf{h}(\mathbf{c})$ (and also of identifiable functions) of a given system $\Sigma$ and it is shown that the corresponding IO-equations $\sys$ can be written in terms of $\mathbf{h}$. 
In this section, we show when the given system $\Sigma$ can be replaced by another realization of $\sys$, say $\Sigma_0$, written only in terms of $\mathbf{h}$ instead of $\mathbf{c}$. 
In this way, $\mathbf{h}$ can be seen as the parameters in $\Sigma_0$ and $\sys$ such that exactly the $\mathbf{h}$ are identifiable. 
As a preparation step, we present an algorithm for finding and evaluating transcendental parameters at suitable values for getting only locally identifiable parameters. The algorithm is an adaptation of~\cite[Algorithm 1]{ilmer2022more}.

\subsection{Preliminaries}
\begin{remark}
Throughout the paper, we have to decide field membership at several steps. 
For a given rational function $f \in \overline{\CC(\mathbf{h})}(\mathbf{x})$, \cite{muller1999basic} has an algorithm for deciding whether $f \in \CC(\mathbf{h})(\mathbf{x})$. 
The problem is that, by making an ansatz for $f$ with undetermined coefficients, we do not yet know how to algorithmically find conditions on these coefficients such that the membership is fulfilled.
\end{remark}

In the following, let us denote $\ksmall:=\CC(\mathbf{h}) \subset \CC(\mathbf{c})$. 
We will look for realizations and parametrizations with coefficient field $\ksmall$ and call them \textit{$\ksmall$-realizations} and \textit{$\ksmall$-parametrizations}, respectively. 
Note that, for polynomially parametrizable curves, the optimal field of rational and polynomial parametrizations coincide~\cite[Section 4, Lemma]{andradas1999base}. 
We will show a comparable result for realizations in Theorem~\ref{thm-optimalrealization}.

Since the parametrization $\Pa$ corresponding to a realization does not involve any additional field extensions, the existence of an $\ksmall$-realization implies the existence of an $\ksmall$-parametrization of the same IO-equation. 
The computation of rational parametrizations of curves over optimal fields is presented in~\cite{sendra2008rational}. 
For the computation of reparametrizations of rational parametrizations of higher-dimensional varieties over optimal fields, we refer to~\cite{andradas2009simplification}.

\subsection{Preparation step: finding locally identifiable reparametrization}\label{sec:aux}
Assume that $\Pa$ is the (proper) rational (or polynomial) parametrization over $\KK=\CC(\mathbf{c})$ corresponding to an observable (polynomial) realization 
\[ \Sigma = \begin{cases}
\mathbf{x}'=\mathbf{f}(u,\mathbf{x}),\\ y=g(u,\mathbf{x})
\end{cases}\]
of $F$. Instead of considering the parametrization $\Pa$ corresponding to $\Sigma$ over $\KK$, we view $\Pa$ over a field generated by the IO-identifiable $\mathbf{h} \in \CC(\mathbf{c})$, which is constructed as follows:
$\{h_1,\ldots,h_\ell\} \subset \CC(c_1,\ldots,c_k)$ is a minimal set of generators of $\CC(h_1,\ldots,h_\ell)$; in particular, $h_1,\ldots,h_\ell$ are algebraically independent.
\begin{itemize}
    \item If $\ell=k$, by the inverse function theorem, $c_1,\ldots,c_k \in \overline{\CC(h_1,\ldots,h_k)}$. So, by the primitive element theorem, there exists $\alpha \in \overline{\CC(h_1,\ldots,h_k)}$ such that $\CC(c_1,\ldots,c_k) = \CC(h_1,\ldots,h_k,\alpha)$.
    \item If $\ell<k$, we use the following iterative construction:
    \begin{itemize}
    \item Set $\ksmall_0 = \ksmall = \CC(h_1,\ldots,h_\ell)$.
    \item For $i>0$, check whether $c_i \in \overline{\ksmall_{i-1}}$. If not, let $\ksmall_i=\ksmall_{i-1}(c_i)$ and    $h_{\ell+i} = c_i$ and otherwise, let $\ksmall_i=\ksmall_{i-1}$  and $h_{\ell+i} = 0$. 
    \end{itemize}
    Then, as above, there is $\alpha$, algebraic over $\ksmall_{k-\ell} = \CC(h_1,\ldots,h_k)$, such that $\CC(c_1,\ldots,c_k) = \CC(h_1,\ldots,h_k,\alpha)$. 
\end{itemize}
Note that, by the rationality of $\mathbf{h}$ in $\mathbf{c}$, renaming variables does not lead to a different result. Moreover, $\ksmall_{k-\ell}$ is a minimal field extension of $\ksmall$ such that $[\KK : \ksmall_{k-\ell}]$ is finite. 

In the above construction, we have to check whether $c_1$ is algebraic over $\ksmall$. 
We do this by computing the intersection ideal
\[ \langle W \cdot H(\mathbf{c})-1, \denom(h_i(\mathbf{c}))\,h_i - \numer(h_i(\mathbf{c})) \mid 1 \le i \le \ell \rangle \cap \CC[h_1,\ldots,h_\ell,c_1] \]
where $H$ denotes the common denominator of $h_1,\ldots,h_\ell$. If there is a polynomial $p \in \CC[h_1,\ldots,h_\ell,c_1] \setminus \CC$ in this intersection ideal, which can be found for instance by Groebner basis computations, then $c_1$ is algebraic over $\ksmall$. Otherwise $c_1$ is transcendental over $\ksmall$. 
For $i>1$ we proceed similarly with adding the transcendental $c_i$'s as new variables $h_{\ell+1},\ldots$ whenever necessary.

Let $F \in \ksmall\{y,u\}$. 
Let $\ksmall_{k-\ell}(\alpha)$ be the coefficient field of a parametrization $\Pa$ of $\Va(F)$, algebraic over $\ksmall_{k-\ell}$ of degree $n$. 
Now we evaluate the transcendental $c_i$ (cf.~\cite{ilmer2022more}), corresponding to the non-zero $h_{\ell+1},\ldots,h_k$, at values $\mathbf{a} \in \ksmall^{k-\ell}$, or even $\mathbf{a} \in \CC^{k-\ell}$, such that the Jacobian matrices $\mathcal{J}(\mathbf{h})$ (the Jacobian with respect to $\mathbf{c}$) and $\mathcal{J}(\Pa)$ have the same rank after the evaluation, and $\Pa$ remains a proper parametrization of $\Va(F)$  if it was proper to begin with. In particular, the evaluation of $\Pa$ at $\mathbf{a}$ is defined, the Jacobian $\mathcal{J}(\Pa)$ evaluated at $\mathbf{a}$ has full rank and the degree does not change. See~\cite[Section 5, Theorem 5.5]{falkensteiner2023rationality} for details. 
Moreover, $F$ is invariant under such evaluation, and its parametrization $\Pa$ corresponds to a realization $\Sigma_{\mathbf{a}}$ of $F$, namely $\Sigma$ evaluated at $\mathbf{a}$.

With the above evaluation, we have found a realization $\Sigma_{\mathbf{a}}$ of $F$  that has coefficients in a field
 algebraic over $\ksmall$, cf.~\cite{OPPS2023,MOS2023}. 
In the following, we want to get rid of the algebraic field extension, which is, in contrast to the transcendental extensions, not always possible. 
In the case of first-order IO-equations, we derive an algorithm for computing an $\ksmall$-realization, if possible.
In the higher-order case, two procedures are presented.

\subsection{Finding globally identifiable reparametrization: first order IO-equations}\label{sec:1storder}
In this section, we assume that the IO-equation $F$ is of order one, i.e., $F \in \ksmall[u,u',y,y']$. 
Assume that the parametrization $\Pa$ corresponds to the observable realization $\Sigma$  whose coefficients generate a field
$\KK$ that is algebraic over $\ksmall$ of degree $n$; say $\KK:=\ksmall(\alpha)$ with $\alpha$ algebraic over $\ksmall$ of degree $n$. 
 Following Section~\ref{sec:simplcoeff}, we consider the  witness variety $\mathcal{H}\subset \overline{\ksmall}^n$ of $\Pa$  (see Definition~\ref{def:witness}). Now, let us  consider Proposition \ref{prop:hypercircles}. If $\mathcal{H}$ has a one-dimensional component parametrizable over $\ksmall$, say $(s_0(z),\ldots,s_n(z))\in \ksmall(z)^n$, it holds that
\[ \mathcal{Q}(z):=\Pa\left(s(z)\right),\quad s(z) :=  \sum_{i=0}^{n-1} s_i(z) \,\alpha^{i}, \]
is a birational parametrization with coefficients in $\ksmall$. 
Thus the corresponding realization
\[\Sigma_0 = \begin{cases}
z'=\dfrac{f(u,s)}{\partial_z(s)},\\ \noalign{\vspace*{1mm}} y=g(u,s).
\end{cases}\]
has coefficients in $\ksmall$. 
Furthermore, if $\mathcal{H}$ does not contain any component being parametrizable over $\ksmall$, Proposition~\ref{prop:hypercircles} ensures that there is no $\ksmall$-realization.

A special case is when $\Sigma$ and thus $\Pa$ is polynomial. 
Then the  witness   variety $\mathcal{H}$ has to define a line  (see Proposition \ref{prop:lines}). 
In the affirmative case, we choose \[(s_0,\ldots,s_{n-1})=\mathbf{p}+z \cdot (\mathbf{p}-\mathbf{q})\] for two points $\mathbf{p},\mathbf{q} \in \mathcal{H}\cap \ksmall^n$ and $s=\sum_{i=0}^{n-1} s_i(z)\,\alpha^{i}$ as before leading again to the parametrization $\mathcal{Q}:=\Pa(s) \in \ksmall[u,z] \times \ksmall[u,u',z]$ corresponding to the polynomial $\ksmall$-realization
\[\Sigma_0 = \begin{cases}
z'=\dfrac{f(u,s)}{\partial_z(s)},\\  \noalign{\vspace*{1mm}} y=g(u,s).
\end{cases}\]
If $\mathcal{H}$ does not define a line, there  are no polynomial $\ksmall$-realizations.

\begin{theorem}\label{thm-optimalrealization}
Let $F \in \ksmall[u,u',y,y']$. If $F$ admits an $\ksmall$-realization and a polynomial realization, then there is a polynomial $\ksmall$-realization.
\end{theorem}
\begin{proof}
Let $\Sigma$ be an $\ksmall$-realization. By~\cite[Theorem 3.10]{falkensteiner2023real}\footnote{In the construction of $R$ in~\cite[Theorem 3.10]{falkensteiner2023real}, no field extension has to be made.}, we may assume that $\Sigma$ is observable. 
Since $F$ admits a polynomial realization, $f$ and $g$ have a common denominator of the form $a \cdot (x-b)^m$ for some $a,b \in \ksmall$, $m \in \NN$. 
Then the reparametrization given by $s=\frac{1+bz}{z}$ leads to a polynomial $\ksmall$-realization (cf. Algorithm~\ref{alg-polynomialrealization}).
\end{proof}

The computation of the  witness  variety and its $\ksmall$-parametrization is fully algorithmic in the case where the associated parametrization corresponds to a curve. 
We refer to the auxiliary algorithm for finding a suitable evaluation of values for the transcendental $c_i$'s such as described in Section~\ref{sec:aux} by {\sc SuitableEvaluation}. 
Then we obtain the following algorithms.

\begin{algorithm}[H]
\caption{OptimalRealization FirstOrder}
\label{alg-optimalrealization}
\begin{algorithmic}[1]
    \REQUIRE An observable realization \[\begin{cases}
    x'=f(u,x), \\y=g(u,x)
    \end{cases}
    \] in $\CC(\mathbf{c},u,u',x)^2$ such that the parameters $\mathbf{c}$ may not be (globally) identifiable.
    \ENSURE A $\CC(\mathbf{h})$-realization of the same IO-equation in identifiable parameters $\mathbf{h} \in \CC(\mathbf{c})$ if it exists.
    \STATE Compute the identifiable functions $\mathbf{h}$ of $\Sigma$ by using~\cite{Ovchinnikov2021}. 
    \STATE Construct the field $\ksmall_n \supseteq \CC(\mathbf{h})$ as described above and view the parametrization $\Pa = (g,\LD_f(g))$ over $\ksmall_n$.
    \STATE Apply the algorithm {\sc SuitableEvaluation} to evaluate the transcendental elements $h_{k+1},\ldots,h_\ell$ at values $\mathbf{a} \in \CC^{k-\ell}$. Denote by $\Pa_{\mathbf{a}}, f_{\mathbf{a}}, g_{\mathbf{a}}$ the resulting expressions.
    \STATE Check whether the witness variety $\mathcal{H} \subset \overline{\ksmall}^n$ associated with $\Pa_{\mathbf{a}}$ is parametrizable over $\ksmall$ and, in the affirmative case, compute a parametrization $(s_0(z),\ldots,s_{n-1}(z))$ over $\ksmall$.
    \STATE Set $s=\sum_{i=0}^{n-1} s_i(z)\,\alpha^{i}$ and output \[ \begin{cases} z'=\dfrac{f_{\mathbf{a}}(u,s)}{\partial_z(s)}, \\ \noalign{\vspace*{1mm}} y=g_{\mathbf{a}}(u,s).\end{cases}\]
\end{algorithmic}
\end{algorithm}

\begin{algorithm}[H]
\caption{OptimalPolynomialRealization FirstOrder}
\label{alg-optimalpolynomialrealization}
\begin{algorithmic}[1]
    \REQUIRE An observable polynomial realization \[\begin{cases}
    x'=f(u,x), \\ y=g(u,x)
    \end{cases}
    \]in ${\CC\langle\mathbf{c}\rangle}[u,u',x]^2$ such that the parameters $\mathbf{c}$ may not be (globally) identifiable.
    \ENSURE A polynomial $\CC(\mathbf{h})$-realization of the same IO-equation in identifiable parameters $\mathbf{h} \in \CC(\mathbf{c})$ if it exists.
    \STATE Apply steps (1)-(3) as in Algorithm~\ref{alg-optimalrealization}. 
    \STATE Check whether the witness variety  $\mathcal{H} \subset \overline{\ksmall}^n$ associated with $\Pa_{\mathbf{a}}$ defines a line over $\ksmall$ and, in the affirmative case, compute a parametrization $(s_0(z),\ldots,s_{n-1}(z))$ over $\ksmall$.
    \STATE Set $s=\sum_{i=0}^{n-1} s_i(z)\,\alpha^{i}$ and output \[ \begin{cases}z'=\dfrac{f_{\mathbf{a}}(u,s)}{\partial_z(s)},\\ \noalign{\vspace*{1mm}} y=g_{\mathbf{a}}(u,s). \end{cases}\]
\end{algorithmic}
\end{algorithm}

Let us note that in the rational and in the polynomial case, the degrees of the right hand sides in the given realization $\Sigma$ and the $\ksmall$-realization $\Sigma_0$ computed by Algorithm~\ref{alg-optimalrealization} and Algorithm~\ref{alg-optimalpolynomialrealization}, respectively, coincide, but their supports might be different.

\begin{example}
Let \[ \Sigma = 
\begin{cases}x'= c(c^2x+1)/3,\\
y=-x^3
\end{cases}\] be a polynomial realization (so, $\KK = \CC(c)$) of the IO-equation \[F=(y'-c^3y)^3 + c^3y^2\] with the identifiable function $h=c^3$, so $\ksmall = \CC(c^3)$. 
The parametrization corresponding to $\Sigma$ can be written as \begin{equation}\label{eq:Pex43}\Pa = \left(-x^3, -(hx + h^{1/3})x^2\right).
\end{equation}
We now introduce $\alpha = h^{1/3}$ and consider the following substitution
\begin{equation}
\label{eq:43subs}
x = z_0+z_1\alpha+z_2\alpha^2,
\end{equation}
which we substitute into~\eqref{eq:Pex43}, expand the result as a polynomial in $\alpha$ and set the coefficients of $\alpha$ and $\alpha^2$ equal $0$ to define the  witness variety $W$
\begin{align*} W=& \mathbb{V}(\{
-3 h z_{0} \,z_{2}^{2}-3 h \,z_{1}^{2} z_{2} -3 z_{0}^{2} z_{1}
, 
-3 h z_{1} \,z_{2}^{2}-3 z_{0}^{2} z_{2} -3 z_{0} \,z_{1}^{2}
,  \\ &\qquad 
-3 h^{2} z_{0} \,z_{2}^{2}-3 h^{2} z_{1}^{2} z_{2} -3 h \,z_{0}^{2} z_{1} -2 h z_{1} z_{2} -z_{0}^{2}
, \\ &\qquad 
-3 h^{2} z_{1} \,z_{2}^{2}-3 h \,z_{0}^{2} z_{2} -3 h z_{0} \,z_{1}^{2}-h \,z_{2}^{2}-2 z_{0} z_{1} \}). \end{align*}
 A computation in {\sc Maple} by using the prime decomposition shows that 
\[ W=\mathbb{V}(\{z_0, z_2\}) \cup \mathbb{V}(\{z_0, z_1, z_2\}),\] 
and hence  
 $W$ consists of the union of the point $(0,0,0)$ and the line parametrized as  $(0,z_1,0)$. Substituting the latter into~\eqref{eq:43subs}, we obtain \[x=z_1\alpha = z_1h^{1/3}.\] This leads to the $\CC(h)$-parametrization \[\mathcal{Q} =
\left(-hz_1^3, -(h^{4/3}z_1 + h^{1/3})z_1^2h^{2/3}\right)=
\left(-hz_1^3,-(h^{2}z_1+h)z_1^2\right)\] corresponding to the $\CC(h)$-realization 
\[ \Sigma_0 = \begin{cases}
z_1'= \dfrac{hz_1+1}{3} ,\\  \noalign{\vspace*{1mm}} y=-hz_1^3 \end{cases}. \]
\end{example}

\subsection{Finding globally identifiable reparametrization: general case}\label{sec:43}
In this section, we present an approach for finding  an $\ksmall$-realization $\Sigma_0$ for a given realization $\Sigma$ of a higher-order IO-equation. This is a direct generalization of the curve-case from Section~\ref{sec:1storder}. We also provide examples illustrating the approach, including converting an only locally identifiable SEIR epidemic model into a globally identifiable one.
\subsubsection{Approach}
So, we consider the case of  IO-equations $\sys$ corresponding to a given ODE system $\Sigma$ in a tuple of state variables $\mathbf{x}$. The IO-equations define an algebraic variety $\Va(\sys)$ of arbitrary dimension. 
Assume that  the coefficients of the parametrization $\Pa$ corresponding to $\Sigma$
 generate a field 
$\ksmall$ that is algebraic over $\KK$ of degree $n$. 

A   characterization in terms of the  witness variety $\mathcal{H}\subset \ksmall^n$ associated to the parametrization $\Pa$ is given by ultraquadrics~\cite{andradas2009simplification}. For a precise definition of ultraquadrics, we refer to  \cite[Definition~16]{andradas2009simplification}. 
It is shown that, whenever $\mathcal{H}$ has an ultraquadric as component, then a reparametrization in $\ksmall = \CC(\mathbf{h})$, where $\mathbf{h}$ are the IO-identifiable functions, can be found. 
An algorithmic method for checking whether the components of $\mathcal{H}$ define an ultraquadric and, in the affirmative case, compute a parametrization of it, however, is missing. 

Nevertheless we will use the weaker shortcut introduced in Section \ref{sec:simplcoeff}. More precisely, if $\Pa$ can be reparametrized over $\KK$ then $\mathcal{H}$ has a component of dimension $\dim(\mathbb{V}(\sys))$ and is (proper) $\KK$-parametrizable. Furthermore, if this component is (proper) $\ksmall$-parametrizable as $(s_0,\ldots,s_{n-1}) \in \ksmall(\mathbf{z})^n$, similarly as in the case of first-order IO-equations, we can use the reparametrization given by $$s=s_0(\mathbf{z})+\cdots+s_{n-1}(\mathbf{z}) \cdot \alpha^{n-1}$$ to obtain the $\ksmall$-realization
\[
\Sigma_0 = \begin{cases}
\mathbf{z}' = \mathcal{J}(\mathbf{s})^{-1} \cdot \mathbf{f}(\mathbf{u},\mathbf{s}), \\
\mathbf{y} = \mathbf{g}(\mathbf{u},\mathbf{s}).
\end{cases}
\]
\subsubsection{Examples}
In this section, we provide examples illustrating our approach. We begin with two technical examples to illustrate the steps of the approach and end with reparametrizing an SEIR epidemiology model to demonstrate a real-life application of the approach.
\begin{example}\label{ex-ultraquadric1}
Let \[\Sigma = \begin{cases}x_1' = \dfrac{x_2^3}{2x_1},\\  \noalign{\vspace*{1mm}}  x_2' = \dfrac{cx_1 + x_2}{3x_2^2},\\ \noalign{\vspace*{1mm}} y = x_1^2\end{cases}\] be a given ODE model. Then $\KK = \mathbb{C}(c)$.
The corresponding IO-equation is $$F=c^6y^3 - 3c^4y^2y''^2 + 3c^2yy''^4 - y''^6 + 6c^2yy'y'' + 2y'y''^3 - y'^2$$ and therefore, the identifiable function is $h(c)=c^2$, and so $\ksmall = \CC(c^2)$. 
The parametrization corresponding to the given realization is 
\begin{equation}\label{eq:44P}\Pa=(x_1^2, x_2^3, cx_1 + x_2).
\end{equation}
Let $\alpha = c = h^{1/2}$ and 
consider  the substitutions
\begin{equation}\label{eq:44xt}
\begin{aligned}
x_1 = z_{1,0} + \alpha z_{1,1},\\
x_2 = z_{2,0} + \alpha z_{2,1}.
\end{aligned}
\end{equation}
Substituting these into~\eqref{eq:44P} and setting the coefficients of $\alpha$ to zero, we obtain the witness variety
\[ W = \mathbb{V}(\{2 z_{1,0} z_{1,1}, h z_{2,1}^{3}+3 z_{2,0}^{2} z_{2,1}, 
z_{1,0}+z_{2,1})
\}.
\]
Decomposing it into irreducible components in {\sc Maple}, we obtain  $W=W_1\cup W_2$ where 
\[ W_1=\mathbb{V}(\{z_{1,0}, z_{2,1}\}), \, W_2=\mathbb{V}(\{z_{1,1}, z_{1,0}+z_{2,1}, h z_{2,1}^{2}+3 z_{2,0}^{2}\}).\]
Since $W_1$ has dimension 2 and $W_2$ dimension 1, we analyze $W_1$. $W_1$ is indeed the plane parametrized properly as 
\[\bm\phi := (0, z_{1,1}, z_{2,0}, 0).\]
Substituting these into~\eqref{eq:44xt}, we obtain the following change of variables
\[
\begin{cases}
x_1 = \alpha z_{1,1}=cz_{1,1},\\
x_2 = z_{2,0},
\end{cases}
\]
and the following new parametrization over $\CC(h)$:
\[
\mathcal{Q}= 
(hz_{1,1}^2, z_{2,0}^3, hz_{1,1} + z_{2,0}).
\]
For simplicity, denote $z_1 = z_{1,1}$ and $z_2 = z_{2,0}$.
From this, we obtain the new realization, now over $\CC(h)$: \[\Sigma_0 =
\begin{cases}z_1' = \dfrac{z_2^3}{2z_1c^2},\\  \noalign{\vspace*{1mm}} z_2' = \dfrac{c^2z_1 + z_2}{3z_2^2},\\   \noalign{\vspace*{1mm}} y = c^2z_1^2,
\end{cases}\] and so $c^2$ can be replaced by the new parameter $h$.
\end{example}

\begin{example}\label{ex-ultraquadric2}
Let \[\Sigma = \begin{cases}x_1' = c x_{2},\\    
x_2' =x_{3}+c ,\\    
x_3' =cx_{1} ,\\
y =c^{2}+x_{1}^{2}
\end{cases}\] 
be a given ODE model. Then $\KK = \CC(c)$. 
The corresponding IO-equation is 
\[F=8 c^{8}-24 c^{6} y +24 c^{4} y^{2}+4 c^{4} y''' -8 c^{2} y^{3}-8 c^{2} y y''' +6 c^{2} y' y'' +4 y^{2} y''' -6 y y' y'' +3 y'^{3}
\] and therefore, the identifiable function is $h(c)=c^2$, and so $\ksmall =\CC(c^2)$. 
The parametrization corresponding to the given realization is 
\begin{equation}\label{eq:45P}\Pa=(c^{2}+x_{1}^{2},\ 2 c x_{2} x_{1},\ 
2 c^{2} x_{2}^{2}+2 c^{2} x_{1}+2 c x_{1} x_{3},\ 
6 x_{2} c^{3}+2 x_{1}^{2} c^{2}+6 c^{2} x_{2} x_{3}
).
\end{equation}
Let $\alpha = c = h^{1/2}$ and 
consider  the substitutions
\begin{equation}\label{eq:45xt}
\begin{aligned}
x_1 = z_{1,0} + \alpha z_{1,1},\\
x_2 = z_{2,0} + \alpha z_{2,1},\\
x_3 = z_{3,0} + \alpha z_{3,1}.
\end{aligned}
\end{equation}
Substituting these into~\eqref{eq:45P} and setting the coefficients of $\alpha$ to zero, we obtain the witness variety
\begin{align*} W  =& \mathbb{V}(2 z_{1,0} z_{1,1}, 
2 h z_{1,1} z_{2,1}+2 z_{1,0} z_{2,0}, \\ &\quad
2 h z_{1,1} z_{3,1}+4 h z_{2,0} z_{2,1}+2 h z_{1,1}+2 z_{1,0} z_{3,0}
,  \\ &\quad
4 h z_{1,0} z_{1,1}+6 h z_{2,0} z_{3,1}+6 h z_{2,1} z_{3,0}+6 h z_{2,0}
).
\end{align*}
Decomposing it into irreducible components in {\sc Maple}, we obtain  $W=W_1\cup W_2\cup W_3$, where 
\[ W_1=\mathbb{V}(z_{1, 0}, z_{2, 1}, z_{3, 1} + 1 ),\ \  W_2=\mathbb{V}(z_{1, 1}, z_{2, 0}, z_{3, 0} ),\ \ W_3=\mathbb{V}( z_{1, 0}, z_{1, 1}, z_{2, 0}, z_{2, 1} ).\]
Moreover, $\dim(W_1)=\dim(W_2)=3$ and $\dim(W_3)=2$. $W_1$ can be parametrized properly as 
\[\bm\phi :=(0, z_{1, 1}, z_{2, 0}, 0, z_{3, 0}, -1).\]
Substituing these into~\eqref{eq:45xt}, we obtain the following change of variables
\[
\begin{cases}
x_1= z_{1,1} \alpha= c z_{1,1} , \\
x_2= z_{2,0}, \\
x_3=-\alpha +z_{3,0}= -c +z_{3,0},
\end{cases}
\] 
and the following new parametrization over $\CC(h)$:
\[
\mathcal{Q}= 
(c^{2} z_{1,1}^{2}+c^{2},\ 2 c^{2} z_{1,1} z_{2,0},\ 
2 c^{2} z_{1,1} z_{3,0}+2 c^{2} z_{2,0}^{2}, \
2 c^{4} z_{1,1}^{2}+6 c^{2} z_{2,0} z_{3,0}
).
\]
For simplicity, denote $z_1 = z_{1,1}$, $z_2 = z_{2,0}$ abnd $z_3=z_{3,0}$.
From this, we obtain the new realization, now over $\CC(h)$: \[\Sigma_0 =
\begin{cases}z_1' = z_{2},\\  \noalign{\vspace*{1mm}} z_2' =  z_{3},\\  \noalign{\vspace*{1mm}} z_3' = c^{2} z_{1},\\ 
\noalign{\vspace*{1mm}} y =c^{2} \left(z_{1}^{2}+1\right),
\end{cases}\] and so $c^2$ can be replaced by the new parameter $h$.
\end{example}

\begin{example}[SEIR model]
Consider the following SEIR epidemic model, which involves
compartments related to disease progression stages with prevalence observations~\cite{SEIR}:
\begin{equation}\label{eq:SEIR}
\begin{cases}
  S' =-\cfrac{bSI}{N},\\
  E' = \cfrac{bSI}{N} - \nu E,\\
  I' = \nu E - a I,\\
  y_1 = I,\\
  y_2 = N.
\end{cases}
\end{equation}
So, we have $\KK = \CC(a,b,\nu, N)$.
The corrersponding IO-equations are:
\begin{gather*}
Ny_1y_1''' + (-Ny_1' + y_1(by_1 + N(a + \nu)))y_1'' - N(a + \nu)y_1'^2 + by_1^2(a + \nu)y_1' + ab\nu y_1^3=0,\\
y_2 - N = 0.
\end{gather*}
Therefore, the IO-identifiable functions are
\[N,\ b,\ a\nu,\ a+\nu,\]
and so $\ksmall = \CC(N,b, a\nu, a+\nu)$.
Denote $h_1 = a+\nu$ and $h_2 = a\nu$.
The parametrization $\Pa$ corresponding to the realization has the following components:
\begin{gather*}
I\\
\nu E-aI\\
\frac{b\nu SI-aN\nu E-N\nu^2E+a^2NI}{N}\\
\frac{bN\nu^2SE-b^2\nu SI^2-2abN\nu SI-bN\nu^2 SI+a^2N^2\nu E+aN^2\nu^2E+N^2\nu^3E-a^3N^2I}{N^2}\\
N
\end{gather*}
Let $\alpha = a$. We have that  $\alpha$ is a root of the polynomial 
\[X^2 - (a+\nu)X + a\nu,\] which has IO-identifiable coefficients.
Since $\alpha$ is algebraic over the field of IO-identifiable functions $\CC(N,b,a+\nu,a\nu)$ of degree $2$, we consider the following change of variables:
\begin{equation}
\label{eq:SEIRchange}
\begin{cases}
S =z_{1,0} + \alpha z_{1,1}\\
E =z_{2,0} + \alpha z_{2,1} \\
I = z_{3,0} + \alpha z_{3,1}
\end{cases}
\end{equation}
Substituting these into $\Pa$ and setting the coefficients of $\alpha$ to zero, we obtain the witness variety as the following union of irreducible varieties (decomposed in {\sc Maple})
\begin{gather*}
W_1 = \mathbb{V}(z_{1, 0}, z_{3, 1}, z_{3, 0} + z_{2, 0} )\\ W_2 = \mathbb{V}(z_{2, 0}, z_{2, 1}, z_{3, 0}, z_{3, 1})
\end{gather*}
A calculation in {\sc Maple} shows that $\dim W_1 =3$ and $\dim W_2= 2$. So, we pick $W_1$, which also has the following proper rational parametrization:
\begin{equation}\label{eq:tparam}
\bm{\phi} = (0,z_{1,1}, -z_{3,0},z_{2,1}, z_{3,0},0 ).
\end{equation}
For simplicity, denote $z_1 = z_{1,1}, z_2= z_{2,1}, z_3 = z_{3,0}$.
With this, substituting~\eqref{eq:tparam} into~\eqref{eq:SEIRchange}, we obtain the final change of variables:
\begin{equation}
\label{eq:SEIRchange-final}
\begin{cases}
S =a z_1\\
E =-z_3 + a z_2 \\
I = z_3,
\end{cases}
\end{equation}
Now~\eqref{eq:SEIRchange-final}
 results in the following parametrization $\mathcal{Q}$ over the field of IO-identifiable functions $\CC(N,b,a+\nu,a\nu) = \CC(N,b,h_1,h_2)$:
\begin{gather*}
z_3,\\ -h_1z_3 + h_2z_2, \\\frac{((-h_1z_2 - z_3)h_2 + h_1^2z_3)N + bh_2z_1z_3}{N},\\ \frac{(-z_2h_2^2 + h_1(h_1z_2 + 2z_3)h_2 - h_1^3z_3)N^2 - 2h_2z_1(h_1z_3 - \frac{h_2z_2}{2})bN - b^2h_2z_1z_3^2}{N^2},\\ N
\end{gather*}
Combining now~\eqref{eq:SEIR} and~\eqref{eq:SEIRchange-final}, we obtain the new reparametrized system over the field of IO-identifiable functions $\CC(N,b,a+\nu,a\nu)$:
\[
\begin{cases}
z_1' = -\cfrac{bz_1I}{N},\\
z_2' = -\cfrac{I(N-bz_1)}{N},\\
I' = a\nu z_2-(a + \nu)I,\\
y_1 = I,\\
y_2 = N.
\end{cases}
\]
\end{example}

\begin{example}[Bilinear model with input]
Consider the model~\cite[Example~1]{LLV}:
\begin{equation}\label{eq:bilinear-input}
\begin{cases}
x_1'= - p_1x_1 + p_2u,\\
x_2'=  - p_3x_2 + p_4u,\\
x_3'=- (p_1+p_3)x_3 + (p_4x_1+p_2x_2)u,\\
y=x_3.
\end{cases}
\end{equation}
We have $\mathbf{x} = (x_1,x_2, x_3)$, $\mathbf{c} = (p_1,p_2,p_3,p_4)$, and $\mathbf{u}=(u)$. 
Computing the IO-equations (turns out to be an order $4$ ODE in $y$), extracting the coefficients, and simplifying the IO-identifiable field generators, we obtain that the globally IO-identifiable functions are \[\mathbf{F} = \mathbb{C}(p_1p_3, p_2p_4, p_1+p_3).\] 
In particular, $p_1$ and $p_3$ are locally but not globally IO-identifiable and neither $p_2$ not $p_4$ are locally identifiable. A calculation also shows that neither $x_1(0)$ nor $x_2(0)$ are locally identifiable, and so the model~\eqref{eq:bilinear-input} is not observable. However, we will still proceed with our procedure to make the parameters locally and even globally identifiable.

Following Section~\ref{sec:aux} (in this example, the parameterization via Lie derivatives is not proper because some of the initial conditions are not locally identifiable),
$k=4$, $\ell = 3$ and $h_1 =p_1p_3$, $
h_2=p_2p_4$,
 $h_3=p_1+p_3$, and so
 $\mathbb{F}_0 = \mathbb{F} = \mathbb{C}(p_1p_3, p_2p_4, p_1+p_3)$.
Then $\mathbb{F}_1 = \mathbb{F}(p_2)$. We can then take $\alpha = p_1$ so that $\mathbb{F}_1(\alpha) = \mathbb{C}(\mathbf{c})$. We do not display the parametrization $\mathcal{P}$ given by the Lie derivatives of~\eqref{eq:bilinear-input} because it is too big. We can now substitute numbers into $p_2$ in~\eqref{eq:bilinear-input} such that 
\[\mathcal{J}(\mathcal{P}) = 
\begin{pmatrix}
0 & 0 & 1\\
p_4u & p_2u & -p_1 - p_3\\
-p_4(2p_1 + p_3)u& -p_2(p_1 + 2p_3)u& (p_1 + p_3)^2\\
p_4(3p_1^2 + 3p_1p_3 + p_3^2)u& p_2(p_1^2 + 3p_1p_3 + 3p_3^2)u& -(p_1 + p_3)^3
\end{pmatrix}
\] is still of full-rank after the substitution. A calculation in {\sc Maple} shows that, if $p_2 \ne 0$, then the substitution results in a full-rank $\mathcal{J}(\mathcal{P})$. We now have
\[
\mathcal{J}(\mathbf{h}) = 
\begin{pmatrix}
p_3 & 0 & p_1 & 0\\
0 & p_4 & 0 & p_2\\
1 & 0 & 1 & 0
\end{pmatrix},
\]
which is also of full rank if $p_2 \ne 0$ (we have that $p_1 - p_3 \ne 0$ because the parameters $p_1$ and $p_3$ do not get substituted because they are locally identifiable). Let us substitute $p_2 = 1$ for simplicity. Since $p_2p_4$ is IO-identifiable, we then are forced to substitute $p_2p_4$ into $p_4$ so that $p_2p_4$ remains unchanged under the substitution.

We now proceed with finding a globally identifiable reparametrization of~\eqref{eq:bilinear-input} following our approach. We have that $\alpha$ is a solution of the following equation:
\[
X^2 - (p_1+p_3)X + p_1p_3 = 0,
\]
which is an irreducible  polynomial of degree $2$ over the field $\mathbf{F}$. Therefore, we consider the following change of variables:
\begin{equation}\label{eq:bilinearalphasubs}
\begin{cases}
x_1 = z_{1,0}+\alpha z_{1,1},\\
x_2 = z_{2,0}+\alpha z_{2,1},\\
x_3 = z_{3,0}+\alpha z_{3,1}.
\end{cases}
\end{equation}
Substituting~\eqref{eq:bilinearalphasubs} into $\mathcal{P}$ and setting the coefficient of $\alpha$ equal zero, after simplifying in {\sc Maple}, we obtain the witness variety
\[
W = \mathbb{V}\left(z_{3, 1},\ p_2p_4z_{1, 1} + z_{2, 1},\ -(p_1+p_3)p_2p_4z_{1, 1} - p_2p_4z_{1, 0} + z_{2, 0}\right),
\]
which has the following parametrization over the field $\mathbf{F}$ of IO-identifiable functions:
\begin{equation}\label{eq:Wparam}
\bm{\phi} = (z_1,\ z_2,\ (p_1+p_3)p_2p_4z_2 + p_2p_4z_1,\  -p_2p_4z_2,\ z_3,\ 0).
\end{equation}
Substituting~\eqref{eq:Wparam} into~\eqref{eq:bilinearalphasubs}, we obtain the final change of variables
\begin{equation}
\begin{cases}
x_1 = z_1 + \alpha z_2,\\
x_2 = p_2p_4(z_1 + (p_1 + p_3)z_2) -p_2p_4\alpha z_2,\\
x_3 = z_3.
\end{cases}
\end{equation}
The resulting globally identifiable reparametrized ODE system is
\[
\begin{cases}
z_1' = p_1p_3z_2+u\\
z_2' = -(p_1+p_3)z_2-z_1,\\
z_3' = (p_2p_4uz_2-z_3)(p_1+p_3)+2p_2p_4uz_1,\\
y = z_3,
\end{cases}
\]
in which the new parameters are $p_1p_3$, $p_1+p_3$, and $p_2p_4$.
\end{example}

\section*{Acknowledgments}
The authors thank Gleb Pogudin for useful discussions. A part of this work was developed during a research visit of S. Falkensteiner to CUNEF University in Madrid. This work was partially supported by the NSF grants CCF-2212460,  DMS-1760448, and DMS-1853650, CUNY grant PSC-CUNY \#65605-00 53, and by the grant PID2020-113192GB-I00 (Mathematical Visualization: Foundations, Algorithms and Applications) from the Spanish MICINN.

\bibliographystyle{elsarticle-num.bst}
\bibliography{biblio}

\end{document}